\documentclass[aps,prl,twocolumn,groupedaddress,floats,showpacs]{revtex4}
\usepackage{latexsym}
\usepackage{dcolumn}
\usepackage[dvips]{graphicx}
\usepackage{amssymb}
\usepackage{graphics}
\usepackage{amsmath}
\usepackage{epsf}

\begin{document}

%\twocolumn[
%\hsize\textwidth\columnwidth\hsize\csname@twocolumnfalse\endcsname
%\draft

\title{
Graphene magnetoresistance in a parallel
magnetic field: Spin polarization effect}
\author{E. H.\ Hwang and S. Das Sarma}
\address{Condensed Matter Theory Center, Department of Physics,
  University of Maryland, College Park, 
Maryland  20742-4111 } 
\date{\today}

\begin{abstract}
We develop a theory for graphene magnetotransport in the presence of
carrier spin polarization as induced, for example, by the application
of an in-plane magnetic field ($B$) parallel to the 2D graphene layer.
We predict a negative magnetoresistance $\sigma \propto B^2$ for
intrinsic graphene, but for extrinsic 
graphene we find a non-monotonic magnetoresistance which is positive at
lower magnetic fields (below the full spin-polarization) and negative
at very high fields (above the full spin-polarization).
The conductivity of  the minority spin band $(-)$ electrons does not vanish
as the minority carrier density ($n_-$) goes to zero. The
residual conductivity of $(-)$ electrons at $n_- = 0$ is unique to
graphene.
We discuss experimental implications of our theory.


\end{abstract}
\vspace{0.5in}
%]

\maketitle

Carrier transport properties of gated 2D graphene monolayers have been
of active current interest
\cite{review,Geim1,Zhang,Fuhrer,Ando,Falko,Hwang1,nomura}, both for
fundamental and technological 
reasons. From a fundamental perspective, understanding the mechanisms
controlling graphene conductivity in experimental samples
is of obvious importance. Such an understanding could lead to
improvement 
in graphene mobilities, which in turn would help its eventual
technological applications. The graphene chiral linear electron-hole
carrier dispersion makes it particularly important that we
understand the operative resistive scattering mechanisms since the
well-understood scattering theories of parabolic 2D
semiconductor structures do not directly apply to graphene because
of its peculiar band dispersion described by the relativistic
Dirac-Weyl equation for massless fermions. In this Letter we predict
that the application of an external magnetic field parallel to the 2D
graphene layer could lead to a novel magnetoresistance behavior in the
low-density and high-mobility graphene samples, providing significant
qualitative insight into its resistive scattering mechanisms.

The physics underlying our predicted novel parallel field graphene
magnetoresistance phenomenon is elegantly simple: The applied field
spin-polarizes the graphene carriers through the Zeeman effect,
dramatically affecting the carrier density of states due to the lifting
of the spin degeneracy, which then leads to a strong modification of
the effective disorder scattering, thereby tuning the system
resistivity. The key point here is that the parallel magnetic field
couples only to the spin degree of freedom since, the 2D graphene layer
being essentially a zero-thickness system, there is no magneto-orbital
coupling to take into account as long as the applied magnetic field is
strictly parallel to the 2D plane. This is different from the situation
in the corresponding parabolic band 2D semiconductor structures
\cite{mr_2d} where both
Zeeman coupling and magneto-orbital coupling \cite{mr_th} play
significant roles in 
determining the parallel field magnetoresistance.

Before describing the details of our theory and results,  we start by
providing a simple physical picture to motivate our predicted
parallel-field-induced graphene magnetoresistance, which could be
either positive or negative depending on the situation. Assuming that
the only effect of the parallel magnetic field is to spin polarize
graphene carriers, we ask how such a continuously tunable (by changing
$B$) carrier spin-polarization would affect the strength of disorder
scattering by modifying the fundamental graphene parameters, namely,
the Fermi wave vector ($k_F$), and the Thomas-Fermi (TF) screening wave
vector ($q_{TF}$), both of which depend on the spin degeneracy factor
$g_s$. Since $k_F \propto g_s^{-1/2}$ and $q_{TF} \propto g_s^{1/2}$,
we note that the Fermi (TF screening) wave vector increases
(decreases) as
the system becomes spin-polarized (with $g_s$ decreasing from 2 to 1)
due to the application of the parallel field. 
Taking into account also the modification of the density of states
($D_F$) due to spin polarization, we discuss below how the tuning of
the carrier spin-polarization in gated graphene will affect the
transport relaxation time, $\tau$, \cite{Hwang1,tau} for various types of
background disorder.

Let us now consider the
various types of resistive scattering potential $V(q)$, which may be
operational in graphene layers. For unscreened short-range white-noise
disorder, $V(q)$ is a constant independent of the wave vector $q$, and
as such, the changing spin-polarization would not affect the
scattering potential, which gives rise to $\tau \propto k_F^{-1}$. In
general, an unscreened long-range disorder 
potential would have $V(q) \sim q^{-n}$ (with $n=1$ being the Coulomb
disorder case for the charged impurity potential). Using $q \sim k_F$,
we conclude that the unscreened long-range disorder would effectively
become weaker in the presence of a finite parallel field since
spin-polarization would in general suppress the disorder strength by
increasing $k_F$, that is, we have $\tau \propto k_F^{2n-1}$. Finally,
the 2D screened Coulomb disorder would have 
$V(q) \sim (k_F + q_{TF})^{-1}$ with $k_F$ ($q_{TF}$) increasing
(decreasing) with the increasing spin polarization under the applied
parallel field, and as such screened Coulomb disorder could either
increase or decrease in the presence of spin-polarization. Remembering
that graphene conductivity in the Boltzmann theory is given by $\sigma
\propto D_F \tau$, where $\tau$ and $D_F\propto \sqrt{g_s}$ are
respectively the 
scattering time and the density of states at the Fermi level we
conclude that if $\tau \propto k_F^{-1}$  (i.e. if the disorder is
independent of spin-polarization), then $\sigma \propto g_s$, and it
will always decrease 
(i.e. a positive magnetoresistance) in the presence of spin
polarization. When the disorder is
spin-polarization dependent (either through the modification of $k_F$
and/or $q_{TF}$), the magnetoresistance could be either positive or
negative. In particular,
for the unscreened long-range disorder potential, $V(q) \sim 1/q^n$, we
have $\sigma 
\propto g_s^{1-n}$. Note that for the bare Coulomb disorder ($n=1$)
$\sigma$ remains a constant and does not depend on the spin
polarization, implying no magnetoresistance.

Below we develop a detailed quantitative theory for the
graphene parallel-field magnetoresistance in the presence of disorder
arising from random charged impurity centers. This is because the
screened charged impurity disorder is almost universally
\cite{Zhang,Fuhrer,Ando,Falko,Hwang1,nomura} 
considered to be the dominant scattering mechanism in
limiting graphene carrier mobility in currently available
samples.

We find that the conductivity of intrinsic (i.e. undoped with
$E_F(B=0)$ precisely at the Dirac point) graphene increases
(i.e. negative magnetoresistance) with
the applied magnetic field. For extrinsic graphene we find
a positive magnetoresistance for $B < B_s$ due to the magnetic field 
induced changes of the screening properties of graphene, where $B_s$
is the field needed for full spin-polarization,
and for $B>B_s$ we find a
negative magnetoresistance due to the depopulation of electrons in the
valence band. This negative magnetoresistance in extrinsic graphene
for $B > B_s$, arising out of magnetic depopulation of the valence
band, is the same qualitative effect which controls the negative
magnetoresistance we find in intrinsic graphene.

%\section{Theory}

We use the Boltzmann transport 
theory  including only the
effect of resistive scattering by random charged impurities 
\cite{Hwang1}. The density of the random 
charged impurity centers is the only unknown parameter in
our model, which  sets the scale of the overall resistivity
without affecting the $\rho(B)$ dependence of
interest in the problem. 
The Boltzmann conductivity of graphene at $T=0$ is given by
$\sigma = {e^2 v_F^2}D_F \tau(E_F)/2$
%\label{eq_sig}
%\end{equation}
where $v_F$ is the Fermi velocity, $D_F = g_s g_v E_F/\gamma^2$
($\gamma = \hbar v_F$) is the 
density of states at Fermi energy ($g_s$ and $g_v$ are spin and valley
degeneracy, respectively), $\tau(E_F) \equiv \tau$ is the
transport relaxation time, and  $E_F = \gamma k_F$ the Fermi energy.
The relaxation time is calculated in the Boltzmann theory as \cite{tau}
\begin{equation}
\frac{1}{\tau(F_F)} = 4\pi n_i r_s^2\frac{v_F}{k_F}
\int_0^1 dx \frac{\sqrt{1-x^2}}{\epsilon(2k_Fx)^2},
\end{equation}
where $k_F$ is the Fermi wave vector, $n_i$ is the background random
charged impurity 
density and $r_s=e^2/\kappa \gamma$ 
($\kappa$ being the background dielectric constant) is the graphene
fine-structure constant, and 
$\epsilon(q)=1 +  v(q) \Pi(q)$ is
the RPA dielectric function of the system, where $v(q) = 2\pi e^2/\kappa
q$ is the  electron-electron 2D Coulomb interaction and 
$\Pi(q)$ the 2D irreducible finite wave vector
polarizability function. 
When the system is spin unpolarized with a 2D carrier density, $n$,
the polarizability function becomes $\Pi(q) =D_F P(q/k_F)$
\cite{Hwang_RPA}, where 
\begin{eqnarray}
P(x) = \left \{ 
 \begin{array}{ll} 1  & \mbox{if $x \leq 2 $} \\
                  1 +\frac{\pi x}{8} - \frac{1}{2}\sqrt{1-
                    \frac{4} {x^2}} 
		  - \frac{x}{4} \sin^{-1}\left ( \frac{2}{x} \right )
                  & \mbox{if $x > 2 $} 
\end{array} 
\right .  .
\end{eqnarray}
Then the conductivity becomes
\begin{equation}
\sigma = \frac{e^2}{h}\frac{n}{n_i}\frac{1}{2r_s^2}\frac{1}{I(2r_s)},
\end{equation}
where $I(x)$ is calculated to be \cite{tau}
\begin{equation}
I(x)  =  \frac{\pi}{4}-\frac{d}{dx}\left [ x^2g(x) \right ] ,
\end{equation}
with 
\begin{eqnarray}
g(x) = \frac{\pi}{2}x-1+\sqrt{|1-x^2|} \left \{
 \begin{array}{cl} {\rm sech}^{-1}({x})
                  & \mbox{if  $x \le 1$} \\
                   \sec^{-1}(x)
                  & \mbox{if $x > 1$} \end{array} 
\right. .
\label{fx}
\end{eqnarray}

When the parallel magnetic field is applied, the
carrier densities $n_{\pm}$ for spin up/down are not equal. Note that 
the total density $n=n_+ + n_-$ is fixed by the external gate. 
The spin-polarized densities
themselves are obtained from the relative shifts (i.e. the
spin-splitting) in the spin up and down bands introduced by the Zeeman
splitting associated with the applied field $B$. 
In this case the polarizability function becomes
\begin{equation}
\Pi(q) =D_F^+ P(q/k_F^+) + D_F^- P(q/k_F^-)
\label{eq_pol}
\end{equation}
where $D_F^{\pm}$ and $k_F^{\pm}$ are the density of
states at Fermi energy and the Fermi wave vector of spin up (down)
state, respectively. 
Note that $g_v$ (=2 in the $B=0$ graphene case) is not lifted, but
the spin degeneracy, by definition, is 
lifted by the in-plane field $B$. The usual unpolarized $B=0$
paramagnetic state has $k_{F}^+=k_{F}^-=k_F$; $n_+=n_-=n/2$.

%%%%%%%%%%%%%%%%%%%%%% FIG. 1 %%%%%%%%%%%
\begin{figure}
\epsfysize=2.2in
\centerline{\epsffile{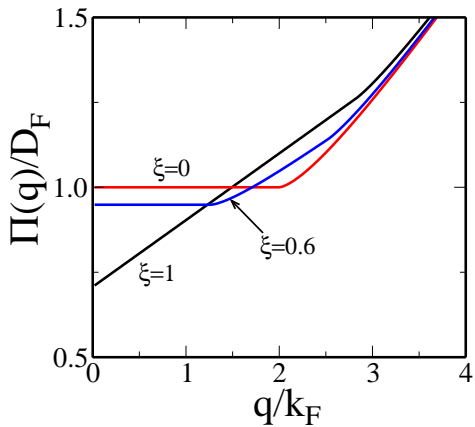}}
%\vspace{0.5cm}
\caption{ Total polarizability of the spin polarized system. Here
  $\xi = (n_+-n_-)/n$, $D_F = g_{v}E_F/\pi\gamma^2$, and $k_F =
  \sqrt{2\pi n/g_{v}}$ is the Fermi wave vector of unpolarized system. 
\label{Fig1}
}
\end{figure}
%%%%%%%%%%%%%%%%%%%%%%%%%%%%%%%%%%%%%%%%%%%%%%%%%%%%%

By defining the spin-polarization parameter given by $\xi =
(n_+-n_-)/n$ we can calculate the irreducible polarizability function in
the case of unequal population of $n_{\pm}$. (Eqs. (1)--(3) above apply
to the $\xi =0$ unpolarized situation.)
In Fig. \ref{Fig1}
we show the irreducible polarizability of the system for different
spin-polarization parameter values.
For unpolarized graphene (i.e. $\xi=0$) we have $n_{\pm} = n/2$ and
$k_F^{\pm} = \sqrt{2\pi n_{\pm}} = \sqrt{\pi n} = k_F$. 
In this case the polarizability $\Pi(q)$ is constant up to $q= 2k_F$
and then increases with $q$.
For fully spin
polarized system (i.e. $\xi=1$) we have $n_+ = n$ and $n_- =0$;
$k_F^+ = \sqrt{2\pi n} = \sqrt{2}k_F$ and $k_F^- =0$.
In this case $\Pi(q)/D_F$ increases linearly up to $q=2k_F^+$ with a slope
$(\pi/8) (k_F/k_F^+) = \pi \sqrt{2}/16$ and has a kink at
$q=2k_F^+$. For the partially polarized system, $0<\xi <1$, $\Pi(q)$
has two distinct 
kink points at $q=2k_F^{\pm}$.

Now we calculate the conductivity in the presence of a parallel magnetic
field for screened Coulomb disorder.
In the presence of the magnetic field
the total conductivity can be expressed as a sum of conductivities of spin
up/down carriers, i.e., $\sigma =\sigma_+ + \sigma_-$,
where $\sigma_{\pm}$ is the conductivity of the ($\pm$) spin
subband, since the two spin
channels are simply parallel transport channels as Coulomb scattering
conserves the spin quantum number and therefore inter-spin-channel
scattering is not allowed. 
The conductivities $\sigma_{\pm}$ are given by 
$\sigma_{\pm} = e^2 v_F^2D_F^{\pm} \tau(E_F^{\pm})/2$.
In calculating the scattering times $\tau(E_F^{\pm})$ we have to use
the two component polarizability given in Eq. (\ref{eq_pol}).
The total carrier resistivity $\rho$ is defined by $\rho \equiv
1/\sigma$.

First, we consider intrinsic
(i.e. undoped or ungated, with $n$ and $E_F$ both being zero) graphene
where the 
conduction band is empty and the valence band fully occupied at
$B=0$. Applying a parallel magnetic field to intrinsic graphene
would cause a shift of the chemical potential from the Dirac point,
and therefore the spin polarized ($+$) electrons induced from the
valence band will
start to occupy the empty conduction band, leaving holes created in
the valence band. (The situation is not dissimilar from what happens
at $T\neq 0$.) The created intrinsic electron density, $n$, is  the
same as the intrinsic hole density, $p$, in the valence band, and the
total density is  
proportional to the applied magnetic field ($B$), i.e.
$n = p=({g_v}/{16 \pi})  ({\Delta}/{\gamma} )^2$, 
%\label{eq_in_n}
%\end{equation}
where $\Delta = g^* \mu_B B$ (with $g^*$ is electron spin $g$-factor
and $\mu_B$ the Bohr magneton) being the Zeeman energy.
Since both carriers contribute to the
conductivity and $n = p$, we can calculate the total conductivity as
$\sigma = \frac{e^2}{h} \frac{n}{n_i}\frac{1}{r_s^2}\frac{1}{I(2r_s)}$.
%\label{eq_in_sig}
%\end{equation}
Since $n \propto B^2$, the calculated conductivity of intrinsic
graphene increases quadratically with the applied magnetic field.  
Note that for this negative magnetoresistance phenomenon in intrinsic
graphene as induced by the applied parallel field, spin-polarization
itself does not play any role -- the physics is dominated by the
field-induced creation of electron-hole pairs which can then carry
current.

In the case of extrinsic graphene (where $E_F, n \neq 0$) an applied
magnetic field produces different numbers of spin up ($+$) and
down ($-$) electrons in the conduction band (or, for holes in the
valence band) and we find 
\begin{equation}
n_{\pm}=\frac{g_v}{4\pi\gamma^2}(\mu \pm \Delta/2)^2
\end{equation}
where $\mu$ is the chemical potential, which
is determined by conserving the total number of
electrons, $n=n_+ + n_-$.  Then we have
$\mu(B) = \gamma \sqrt{\pi n} \sqrt{1-\delta^2/4}$,
where $\delta = {\Delta}/E_F$ and $E_F = \gamma k_F = \gamma
\sqrt{\pi   n}$ is the Fermi energy at $B=0$ (unpolarized system).
When $B=0$ ($\Delta=0$) we have $\mu = E_F$.
We can relate the spin-polarization parameter $\xi$ to the applied
magnetic field as $\xi = \delta \sqrt{1-\delta^2/4}$.
At $\delta = \sqrt{2}$ the system is
fully spin polarized (i.e. $\xi=1$). 
Thus, the so-called saturation (or the spin-polarization) field for complete
spin-polarization is 
given by $B_s = \sqrt{2}E_F/(g^*\mu_B)$, which is smaller by a factor
of $\sqrt{2}$ than
$B_s = 2 E_F/(g^*\mu_B)$ for a 2D parabolic band dispersion.
The conductivity of the partially polarized system can be calculated as
\begin{equation}
\sigma = \frac{e^2}{h} \frac{1}{n_i}\frac{1}{2r_s^2}\left [ \frac{n_+}{I_+} +
  \frac{n_-}{I_-} \right ],
\end{equation}
where
\begin{equation}
I_{\pm} = \int_0^1 dx \frac{\sqrt{1-x^2}}{\varepsilon(2k_F^{\pm}x)^2}.
\end{equation}

%%%%%%%%%%%%%%%%%%%%%% FIG. 2 %%%%%%%%%%%
\begin{figure}
\epsfysize=2.2in
\centerline{\epsffile{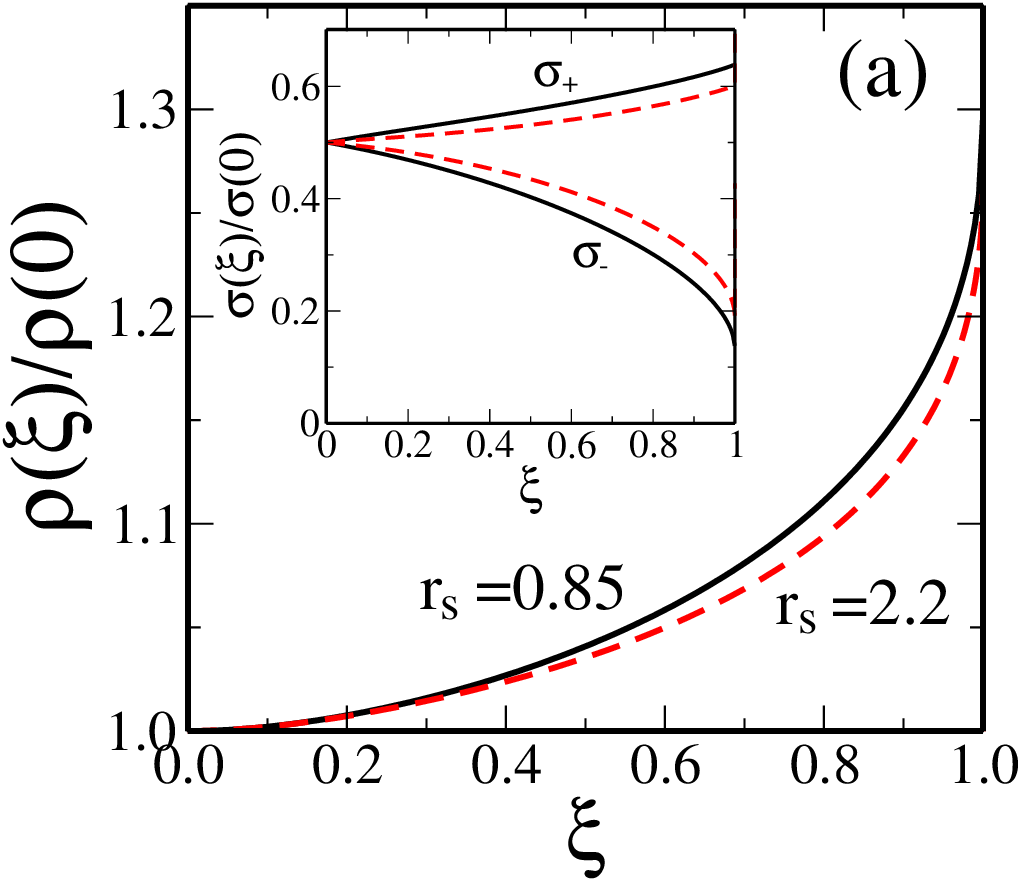}}
\epsfysize=2.2in
\centerline{\epsffile{Fig2b.eps}}
%\vspace{0.5cm}
\caption{(a) Calculated resistivity $\rho(\xi)/\rho(0)$ as a function of the
  spin polarization for $r_s=0.85$ (solid lines) and 2.2 (dashed lines). Here 
  $\rho(0)$ is the resistance at $B=0$.
Inset shows the conductivities of each spin state.  
Note that the conductivity $\sigma_-$ does not vanish at $\xi = 1$ (or
$B=B_s$).
(b) Magnetoresistance at $B_s$, $\rho(B_s)/\rho(0)$, as a function
  of interaction parameter $r_s$. (Note that $r_s >2.2$ is not a
  physical regime.)
\label{Fig2}
}
\end{figure}
%%%%%%%%%%%%%%%%%%%%%%%%%%%%%%%%%%%%%%%%%%%%%%%%%%%%%

In Fig. \ref{Fig2} we show the calculated resistivity, $\rho =
1/\sigma$, as a 
function of spin polarization for $r_s = 0.85$ and 2.2 which
correspond to graphene on SiO$_2$ substrate and vacuum. Note that we
can represent the Fermi wave vectors as a function of the spin
polarization, $k_F^{\pm} = k_F \sqrt{1\pm 
  \xi}$. The total resistivity  increases as the spin polarization (or
magnetic field) increases because of overall suppression of screening.
At $\xi = 1$  (or $B=B_s$) we have
\begin{equation}
I_+ = \frac{1}{\left ( 1+ \pi r_s/4
  \right )^2}I(\frac{r_s}{1+\pi r_s/4}),
\end{equation}
and as $\xi \rightarrow 1$
\begin{equation}
I_- \rightarrow \frac{\pi}{16}\frac{1}{r_s^2}\frac{n_-}{n_+}.
\end{equation}
Thus, as $\xi \rightarrow 1$, $n_- \rightarrow 0$ and the scattering time
of spin $(-)$ electrons diverges as $1/n_-$ because $\tau_{\pm}
\propto 1/I_{\pm}$. This means that the
conductivity of spin $(-)$ electrons, $\sigma_- \propto n_-/I_-$,
is finite at $\xi =1$ even 
though there is no spin down carrier. 
That is, we have $\sigma_-(B_s)=\frac{e^2}{h}\frac{n}{n_i}
\frac{8}{\pi}$, which is independent of the interaction parameter $r_s$.
This is a consequence of vanishing
density of states of graphene as $n \rightarrow 0$ and the peculiarity
of graphene screening properties, i.e. the contribution of
interband transition to screening.
We also show as an inset in Fig. \ref{Fig2} the calculated
conductivities of each spin subband. As 
expected  $\sigma_-$ does not vanish as $\xi \rightarrow 1$. 
The result in Fig. \ref{Fig2}(a) does not depend on the carrier
density at $B=0$. 
However, it is weakly dependent on the interaction parameter $r_s$.
The maximum magnetoresistance at $B_s$ can be calculated as 
\begin{equation}
\frac{\rho(0)}{\rho(B_s)} =
\frac{\sigma(B_s)}{\sigma(0)} = \frac{I_0}{I_+} + \frac{16 r_s^2}{\pi} 
I_0,
\label{eq_rat}
\end{equation}
where $I_0 = I(2r_s)$, and is shown in Fig. \ref{Fig2}(b). As $r_s
\rightarrow 
0$ we have $I_0,I_+ 
\rightarrow \pi/4$, and the ratio becomes
${\rho(B_s)}/{\rho(0)} \rightarrow 1$.
As $r_s \rightarrow \infty$ $I_0 \rightarrow \pi/64 r_s^2$ and $I_+
\rightarrow (16/\pi^2 r_s^2) I(4/\pi)$, and the ratio becomes
${\rho(B_s)}/{\rho(0)} \approx 1.22$.
%\begin{equation}
%\frac{\rho(B_s)}{\rho(0)} \rightarrow \frac{\pi^3}{1024}
%\frac{1}{I(\frac{4}{\pi})} +\frac{1}{4} \approx 1.22.
%\end{equation}
Thus we always have positive magnetoresistance in extrinsic graphene
as the applied in-plane 
magnetic field increases from 0 to $B_s$.
Note also that $\rho(B_s)/\rho(0)$ has a shallow maximum at $r_s
\approx 0.5$.

\begin{figure}
\epsfysize=2.2in
\centerline{\epsffile{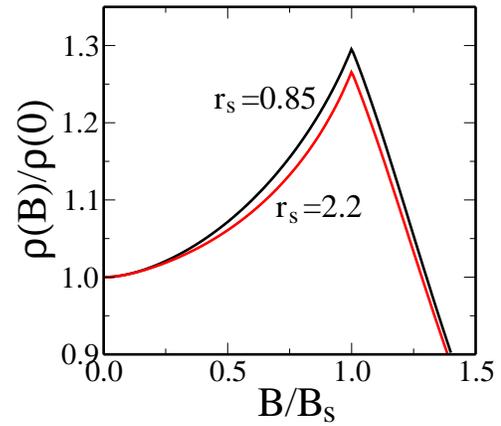}}
%\vspace{0.5cm}
\caption{Calculated resistivity $\rho(B)/\rho(0)$ of graphene
as a function of the in-plane magnetic field for $r_s=0.85$ and 2.2.
\label{Fig3}
}
\end{figure}

As the magnetic field increases beyond $B_s$, i.e. $B>B_s$, the number
density of spin ($+$) state is given by 
\begin{equation}
n_+ = n + p = n + \frac{n}{2}\frac{1}{\delta^2} \left (
  \frac{\delta^2}{2} -1 \right )^2,
\end{equation}
where $p$ is the induced hole density created in the valence band due
to the magnetic field. 
In Fig. \ref{Fig3} we show the calculated resistivity of graphene
as a function 
of magnetic field. A very interesting feature is the sharp decrease of
resistivity 
for $B>B_s$, which is unexpected, when compared with
ordinary 2D systems where the resistivity saturates (or increases) for
$B>B_s$ \cite{mr_2d,mr_th}. 
In graphene, however, for $B \gg B_s$ the calculated
conductivity increases as $B^2$.
This increasing conductivity for $B > B_s$
in extrinsic graphene has exactly the same physical origin as the
negative magnetoresistance we find for intrinsic graphene.
%(c.f. Eq. (\ref{eq_in_sig}) above).

%%%%%%%%%%%%%%%%%%%%%% FIG. 3 %%%%%%%%%%%
%%%%%%%%%%%%%%%%%%%%%%%%%%%%%%%%%%%%%%%%%%%%%%%%%%%%%

Given the nontrivial (and surprising) magnetoresistive behavior of
graphene we predict in a parallel magnetic field, namely, negative
magnetoresistance ($\sigma(B) \sim B^2$) in intrinsic graphene and
positive magnetoresistance ($\rho(B \ll B_s) \sim B^2$) as well as
negative magnetoresistance ($\sigma (B \gg B_s) \sim B^2$) for
extrinsic graphene, a question naturally arises about the experimental
observability of our predicted phenomenon. We believe that our
predictions should be observable at low temperatures ($<1K$) in
high-mobility graphene samples provided reasonable values of
spin-polarization (e.g. $\xi=0.5$ or above) are achieved. Using $g^*
=2 $ for graphene we find $\Delta \approx 0.12 B$ meV where $B$ is
measured in tesla. This leads to $B_s(T) \approx 140 \sqrt{\tilde{n}}$
where $\tilde{n}$ is the graphene carrier density measured in unit of
$10^{10}$ cm$^{-2}$. Thus, for $n = 10^8$ cm$^{-2}$, $B_s \approx
14T$, which should enable a direct verification of our theory close to
the charge neutrality point. We anticipate $g^*$ to be enhanced by
electron-electron interaction effects, perhaps as much as by a factor
of 2 for $r_s =2.2$, and therefore the necessary magnetic field values
may be a factor 2 lower for free-standing graphene in vacuum than on a
substrate. It seems that a verification (or falsification) of our
predicted magnetoresistance behavior should be possible in
high-mobility suspended graphene samples at low carrier densities
\cite{Bolotin}. This would 
help our understanding of the resistive mechanisms and the nature of
disorder in graphene environment.

This work is supported by the U.S.-ONR.

\end{document}